\documentclass[pdflatex,sn-mathphys]{sn-jnl}

\jyear{2022}%

\theoremstyle{thmstyleone}%
\theoremstyle{thmstyletwo}%

\theoremstyle{thmstylethree}%

\begin{document}

\title[ ]{Forensic Dental Age Estimation Using Modified Deep Learning Neural Network}


\author*[1]{\fnm{İsa} \sur{Ataş}}\email{isa\_atas@dicle.edu.tr}
\author[2]{\fnm{Cüneyt} \sur{Özdemir}}
\author[2]{\fnm{Musa} \sur{Ataş}}
\author[2]{\fnm{Yahya} \sur{Doğan}}

\affil*[1]{\orgdiv{Computer Technology Department}, \orgaddress{\street{Diyarbakır Vocational School of Technical Sciences}, \orgname{Dicle University}, \city{Diyarbakır}, \country{Turkey}}}

\affil[2]{\orgdiv{Computer Engineering}, \orgname{Engineering Faculty}, \orgaddress{\street{Siirt University}, \city{Siirt}, \country{Turkey}}}


\abstract{Dental age is one of the most reliable methods to identify an individual's age. By using dental panoramic radiography (DPR) images, physicians and pathologists in forensic sciences try to establish the chronological age of individuals with no valid legal records or registered patients. The current methods in practice demand intensive labor, time, and qualified experts. The development of deep learning algorithms in the field of medical image processing has improved the sensitivity of predicting truth values while reducing the processing speed of imaging time. This study proposed an automated approach to estimate the forensic ages of individuals ranging in age from 8 to 68 using 1332 DPR images. Initially, experimental analyses were performed with the transfer learning-based models, including InceptionV3, DenseNet201, EfficientNetB4, MobileNetV2, VGG16, and ResNet50V2; and accordingly, the best-performing model, InceptionV3, was modified, and a new neural network model was developed. Reducing the number of the parameters already available in the developed model architecture resulted in a faster and more accurate dental age estimation. The performance metrics of the results attained were as follows: mean absolute error (MAE) was 3.13, root mean square error (RMSE) was 4.77, and correlation coefficient R$^2$ was 87\%. It is conceivable to propose the new model as potentially dependable and practical ancillary equipment in forensic sciences and dental medicine.}




\maketitle

\section{Introduction}\label{sec1}

Age determination is a critical research topic in forensic science \cite{1}. Age range based on facial features is considered the most common approach due to its applicability \cite{2}. In forensic units, in cases where the person's age is uncertain and disputed, a medical age assessment may be requested by government agencies for legal action. Therefore, the age estimation process may provide guidance on whether an individual should be considered a child, teenager, or adult \cite{3}. Despite variances in many countries, the legal age range for criminal liability is typically between 14 and 21 \cite{4, 5, 6}. The radiographic images of hand bones (metacarpus) and wrist bones (pisiform) and the DPR images are utilized frequently in forensic sciences to identify the age range \cite{7}.

This study employed the DPR images for age range determination. After puberty, tooth development decelerates in individuals, and their distinctive tooth properties deteriorate. Therefore, age determination from DPR images becomes challenging after a particular age range \cite{8}. For dental age determination, traditional manual processes are commonly used \cite{9}. Although these manual techniques have been employed properly in diverse populations, there are still certain limitations in clinical applications, such as the technique's subjectivity and measurement bias. These procedures are also monotonous and time-consuming \cite{10}. Therefore, estimating the automatic dental age is crucial to improve the age range accuracy \cite{7}.

Traditional automated dental age estimation procedures involve phases, such as image-preprocessing, segmentation, feature extraction, and classification (categorical) or regression (numerical). In the case of classification, these procedures seek to determine individuals' age groups, whereas the regression phase seeks to pinpoint their precise ages. The success of each step in the methods highly depends on its compatibility with the previous ones \cite{11}.

Deep learning approaches (artificial vision, object detection, recognition, etc.), which have recently supplanted traditional methods, efficiently address numerous crises in diverse scientific fields \cite{12,13,14,15,16}. The deep learning approach is a machine learning process in which deep neural networks such as convolutional neural networks can work directly on the input images and generate the required output without demanding the execution of intermediate steps such as segmentation and feature extraction. However, efforts to design and train deep neural networks are complex, time-consuming, and costly. Therefore, instead of designing and training deep networks from scratch, several approaches were developed that use pre-trained deep networks to execute the necessary tasks. Transfer learning is the common term for these methods \cite{17}. Additionally, this study adopted the transfer learning approach in the deep learning model proposed for forensic age estimation. The mentioned model aims to accurately predict an individual's forensic age from DPR images in a precise range.

The proposed approach consists of three steps: image-preprocessing, feature extraction, and linear regression. Initially, DPR images are preprocessed and prepared for feature extraction in the image-preprocessing step. The data augmentation method is employed on the data set for this function. In the feature extraction step, however, the feature extraction capability of the pre-trained InceptionV3, ResNet50V2, DenseNet201, MobileNetV2, VGG16, and EfficientNetB2 deep-learning models were applied. Finally, age estimation was made using statistical methods (machine learning algorithms) in the stage of linear regression.

The remainder of this article is structured as follows: Section 2 includes similar and related works. Section 3 describes materials and methods, and section 4 presents experimental analysis results. The last section, concludes the final evaluations and future perspectives.

\section{Related Works}
\label{sec:2}
 
Čular et al. \cite{7} proposed a dental age estimation method based on DPR images. Accordingly, they combined active appearance and active shape models to determine the outer contour of the teeth. They employed statistical models for feature extraction and a neural network model for age estimation. They generated a dataset including DPR images of 203 individuals to test the effectiveness of the suggested strategy. The data set revealed that the active shape and appearance models had mean absolute errors (MAE) of 2.481 and 2.483, respectively. 

Štepanovský et al. \cite{18} compared the performance of 22 different age estimation approaches in terms of their accuracy and complexity in estimating dental age. They used a dataset containing 976 DPR images. Experimental results suggested that the best practices were multiple linear regression models, tree models, and support vector machine (SVM) models. 

Hemalatha et al. \cite{19} proposed a classification model grounding on the Demirjian approach to estimate the dental age of Indian children. They used a dataset of images from 100 healthy individuals aged between 4 and 18 to assess the viability of the proposed model. After beginning with preprocessing for noise removal and smoothing from the input images, they segmented the teeth and extracted various features from them. Finally, they applied a fuzzy neural network to perform the classification and concluded that the proposed approach had 89\% accuracy. 

Sironi et al. \cite{1} suggested an age estimation approach based on measuring the amount of the odonblast (pulpal tissue), capturing the images of the pulpal tissue through 3D cone beam computed tomography (CBCT). To estimate dental age in their proposed approach, they generated a dataset containing data on 286 healthy individuals and analyzed it using a Bayesian network. Their findings emphasized that the proposed method was promising for accuracy, bias, and sensitivity from the error matrix (confusion matrix) criteria.

Tao et al. \cite{20} proposed a method of dental age estimation using the multilayer sensor algorithm, one of the machine learning algorithms. They used the k-fold cross-check in the training process to address the overfitting problem. In their experiments, they employed a dataset including 1,636 images. Their findings demonstrated that, in terms of MAE, MSE, and RMSE, the proposed method outperformed the traditional methods such as Demirjian and Willem. 

De Back et al. \cite{21} presented a dental age estimation technique based on DPR images. They employed a Bayesian convolutional neural network for uncertainty and age estimation, conducting experiments on the DPR image dataset of 12,000 individuals. According to the outcomes, they reported that the proposed approach had a correlation coefficient of 0:91. 

Kim et al. \cite{22} defined a deep learning-based automatic dental age estimation approach, using a convolutional neural network (CNN) for age estimation. Their dataset contained 9,435 DPR images from three different age categories. Based on their result, they reported that the proposed approach revealed excellent performance. 

Asif et al. \cite{23} studied a statistical model of dental age estimation Based on a volumetric examination of the dental pulp/tooth ratio. They applied simple linear regression and Pearson correlation analysis to the proposed model. Their dataset included 3D CBCT images of 300 individuals separated into five age groups. Their results indicated that the proposed approach figured a 6.48 MAE. 

Farhadian et al. \cite{24} proposed a dental age estimation approach using a neural network model to perform the age estimation task. They experimentally verified the superiority of the neural network over a regression model using a dataset of 300 CBCT scans in their experiment. They identified the performance metrics as 4.12 MAE and 4.4 RMSE. 

N. Mualla et al. \cite{25} described a transfer learning-based automated method for estimating dental age. Two deep neural networks, AlexNet and ResNet, were used in their strategy to extract features. They used several classifiers to execute the classification task, including a decision tree, k-nearest neighbor, linear discriminant, and support vector machine. They constructed a dataset containing 1,429 DPR images and tested various suitable performance metrics. 

Guo, Yc et al. \cite{10} compared the performance of manual and machine learning methods for age estimation from DPR images. They executed analyzes of a dataset including 10,257 individuals' images. As a result, they reported that the end-to-end CNN models performed better, according to the comparison of the manual method and machine learning.

\section{Materials and Methods}
\label{sec:3}
An automated dental age estimation approach based on DPR images was proposed in this section. This approach sought to identify the age group using DPR images and included three primary steps: image-preprocessing, feature extraction, and regression.
\subsection{Image pre-processing}
\label{sec:3.1}
\begin{figure}[ht]
\centering
\includegraphics[width=0.75\linewidth]{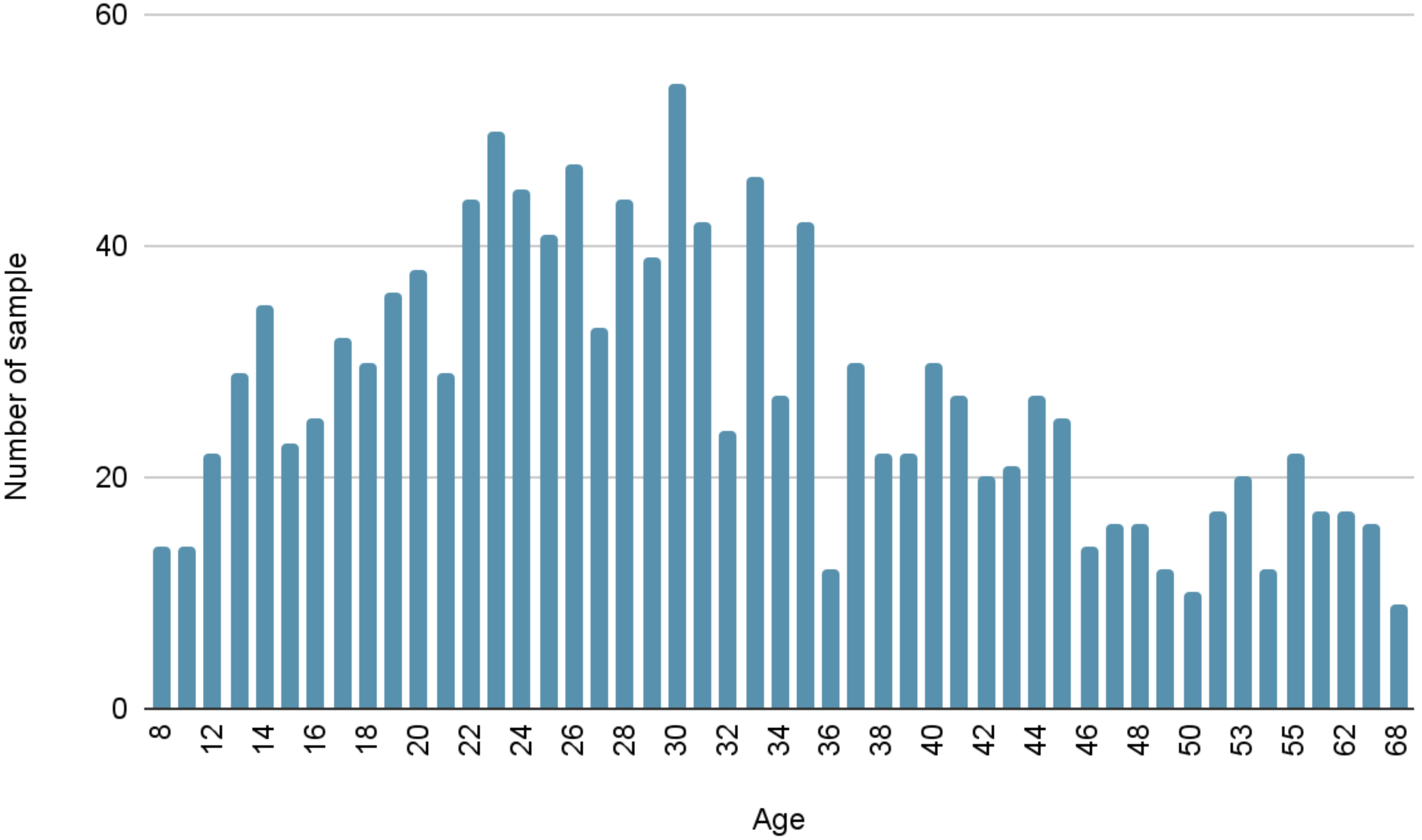}
\caption{Distribution of DPR images by age range}
\label{fig:1}
\end{figure}

\begin{figure}[ht]
\centering
\includegraphics[width=0.90\linewidth]{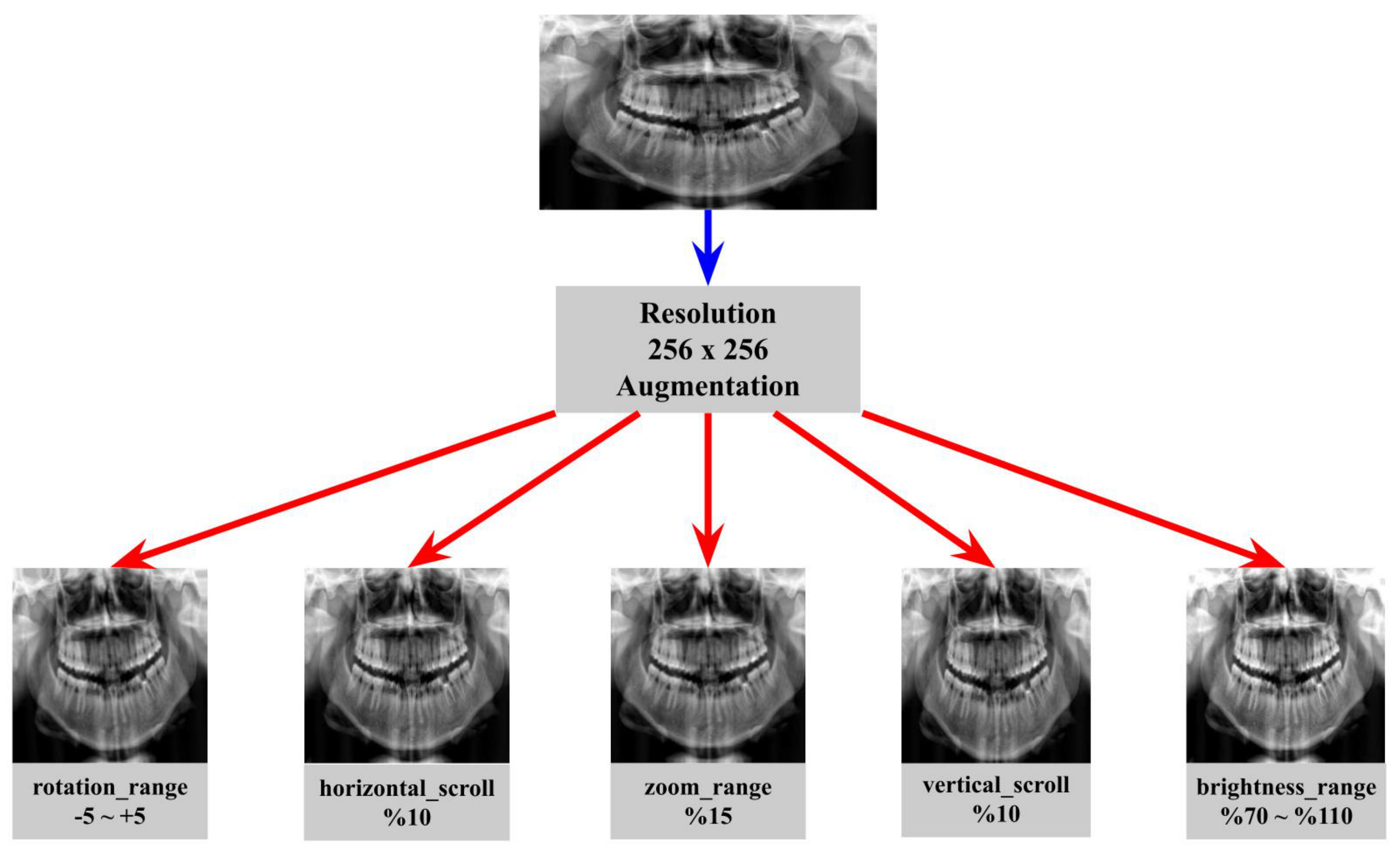}
\caption{Data augmentation process for DPR images}
\label{fig:2}
\end{figure}

A data set that included 1,332 DPR images between the ages of 8-68 and acquired from the Periodontology clinic of Diyarbakır Oral and Dental Health Hospital was analyzed. Fig. \ref{fig:1} depicts the distribution of DPR images by age.
Before processing the feature extraction, the DPR images in the dataset underwent two preprocessing stages. Initially, the images were scaled to 256 × 256 resolution to suit the deep neural network requirements. Then, the data augmentation technique was applied to improve the model's success on diverse images and minimize the overfitting effect. This technique is used in expanding the training data set by entraining additional images that are variants of the existing images in the dataset. In an effort to establish the optimal image values and investigate how the data diversity method's various functions affect images, the fairest values were designated. The DPR images were subjected to the data augmentation technique with the following preferences: rotation between -5 and +5, 15\% augmentation, 10\% horizontal and vertical scrolling, and 70\% - 110\% brightness. Fig. \ref{fig:2} demonstrates the data augmentation for DPR images.
Experimental analyzes were performed on a computer with an Intel® Core™ i7 2.4Ghz CPU @ 1.60 GHz processor with 16GB primary memory, 4-GB NVIDIA GeForce GTX 1080 graphics card.

\subsection{Feature extraction}
\label{sec:3.2}
\begin{figure}[ht]
\centering
\includegraphics[width=\linewidth]{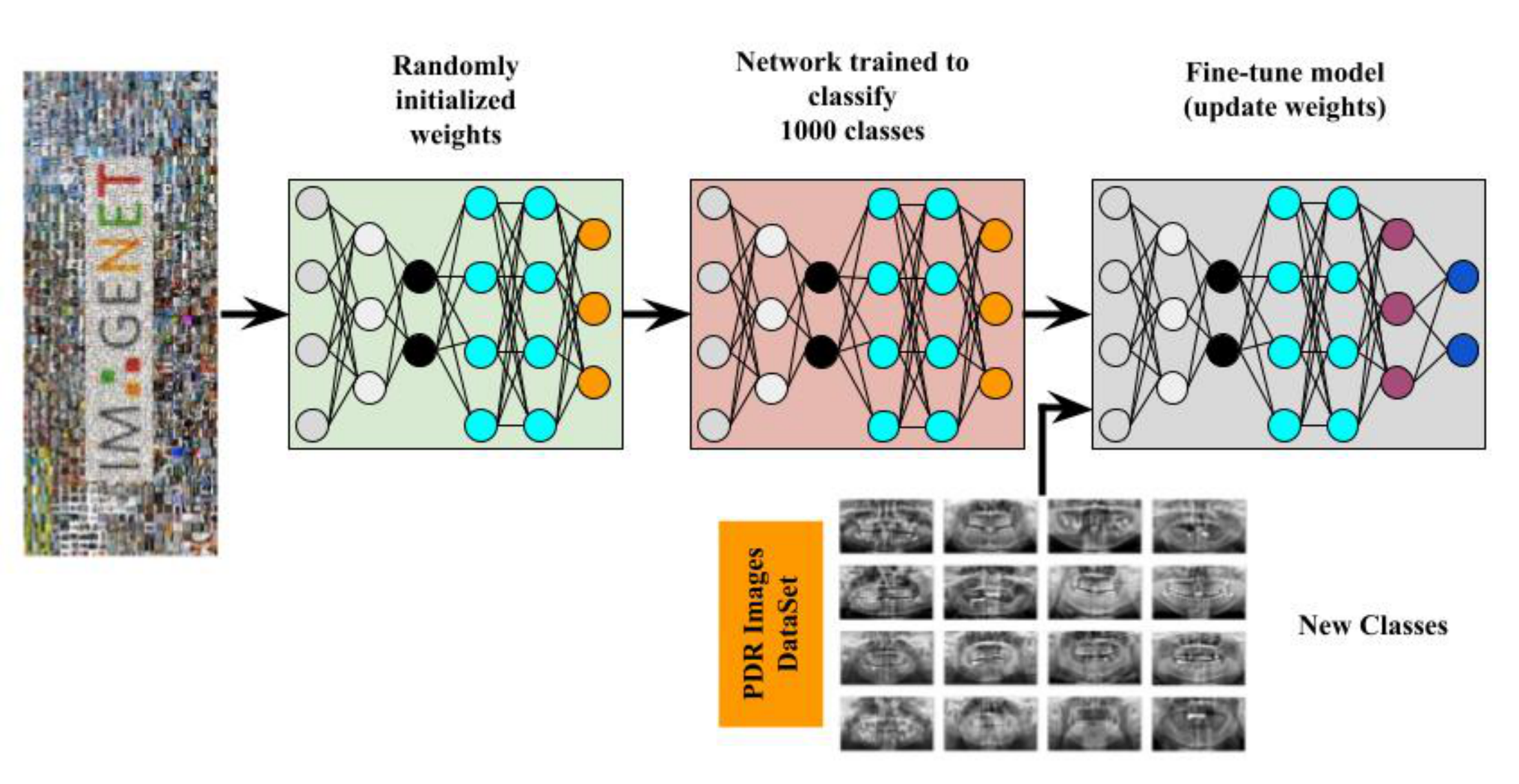}
\caption{Transfer learning procedure}
\label{fig:3}
\end{figure}

The dimensionality reduction procedure known as feature extraction divides the dataset into more logical groups for processing. Convolutional Neural Networks (CNN) is a deep learning algorithm used in image processing and working with images as input. This algorithm utilizes various techniques to capture the images' features and consists of several layers. A CNN typically contains three layers: convolutional, pooling, and fully connected \cite{26}. The proposed study carried out the feature extraction step by applying transfer learning technique to InceptionV3 \cite{27}, MobileNetV2 \cite{28}, ResNet50V2 \cite{29}, EfficientNetB4 \cite{30}, VGG16 \cite{31} and DenseNet201 \cite{32} deep learning models. As a deep learning strategy, transfer learning uses pre-trained model parameters on a sizable dataset (ImageNet, COCO, etc.). In addition, transfer learning is used for many reasons. For instance, it is a challenging issue to train a CNN from scratch using random initial values in case of an insufficient dataset. Therefore, using the weights of a pre-trained network as initial values may address many existing problems effectively. Transfer learning models are used to accelerate the speed of this process and generate the best practicing model. Fig. \ref{fig:3} displays the transfer learning procedure. 
As taken reference in this study, the analyses made with the deep learning models revealed that the InceptionV3 model outperformed other models in terms of performance. As a result, the InceptionV3 model underwent a modification process.

\subsubsection{Feature extraction using InceptionV3}
\label{sec:3.2.1}
\begin{figure}[ht]
\centering
\includegraphics[width=\linewidth]{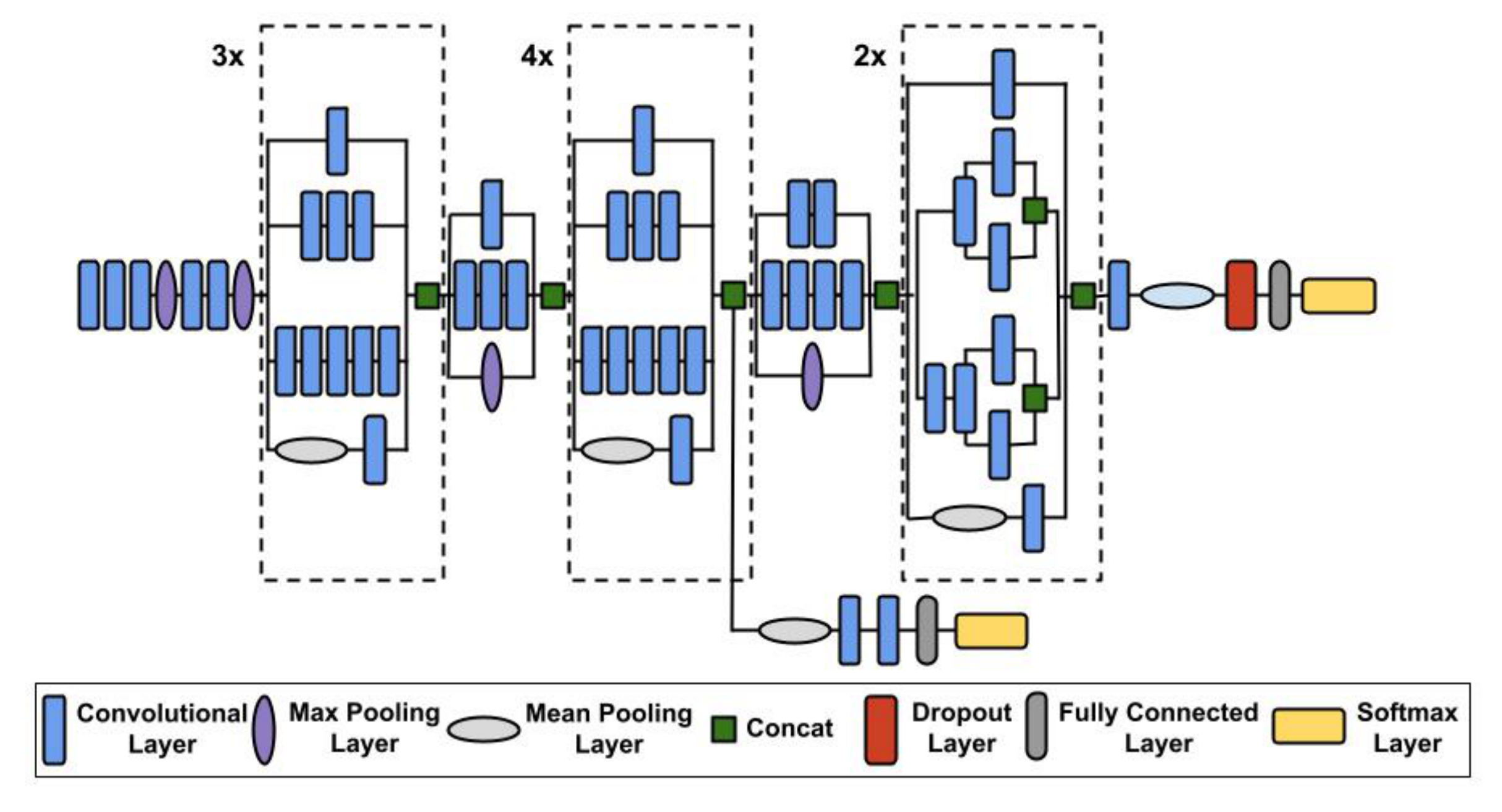}
\caption{The architecture of InceptionV3 model}
\label{fig:4}
\end{figure}

\begin{figure}[ht]
\centering
\includegraphics[width=\linewidth]{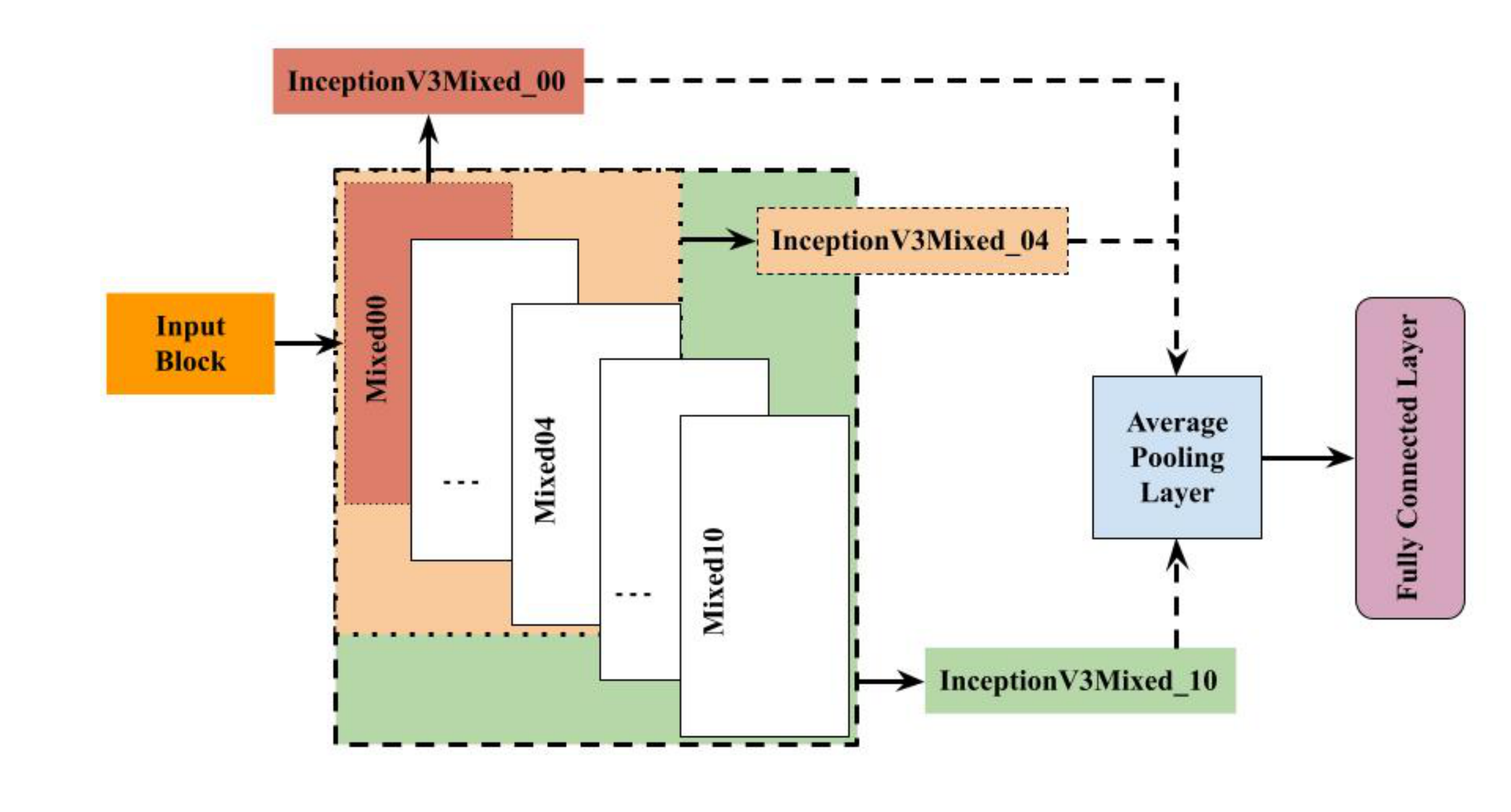}
\caption{Inter-block exchange in InceptionV3 architecture}
\label{fig:5}
\end{figure}

InceptionV3 is a convolutional neural network model used in image analysis and recognition problems and consists of multiple-convolution and max-pooling layers. The bottom layer contains a fully connected neural network. The prominent component of the model is that it replaces small kernels with large kernels by learning multi-scale representations to minimize computational complexity and the total number of parameters \cite{27}. Fig. \ref{fig:4} depicts the typical functional architecture of the InceptionV3 model.
The InceptionV3 model contains 11 different blocks. The model's architecture underwent special modifications/adjustments to achieve higher model success. The blocks labeled 'mixed' were omitted from the model to simplify the algorithm, lower the number of parameters, and obtain the most optimal results from the available data set. As a result, the InceptionV3 model generated a large number of sub-patterns. Subsequently, the generated sub-patterns were labeled with the block number. For instance, the sub-pattern consisting of the first seven blocks was called InceptionV3Mixed7. The InceptionV3Mixed\_04 model performed the best among the sub-patterns. Fig. \ref{fig:5} displays the sub-pattern extraction derived from the InceptionV3 model.

\subsection{Linear Regression }
\label{sec:3.3}
Machine learning refers to the algorithms that train to predict an outcome after learning from a set of user-generated data. The model training proceeds until the performance hits up to the optimal level. An outcome-predicting algorithm from an available input data set is called a supervised learning algorithm, and it involves classification and regression methods to develop prediction models. In a supervised learning algorithm, if the predicted responses are labels, the classification method is used; however, if they are within a particular value range, then the regression method is used. The term regression denotes the analysis of estimating the independent variable from the dependent variables. The utmost preferred two regression types in machine learning are logistic and linear regressions. If the target (independent variable) is categorical, it is typically binary (logistics) regression, whereas the reference is the linear regression if it is continuous. Mean square error (MSE), root mean square error (RMSE), mean absolute error (MAE), and coefficient of determination (R$^2$) are the metrics used for assessing the regression model in convolutional neural network and machine learning architectures \cite{33}.

Mean Square Error (MSE): it measures the performance of the imputed values in the machine learning model, and the results are positive. Metric values close to zero refer to better performance \cite{13}.
\begin{equation}
MSE = \frac{1}{N} \sum_{i=1}^{n} (\widehat{Y_{i}} - Y_{i}) ^{2} \end{equation}

Root Mean Square Error (RMSE): It denotes the square root of the MSE. When the MSE value is significantly larger than the other metrics, it is typically preferred over the MSE metric to facilitate interpretation \cite{13}.
\begin{equation}
RMSE = \sqrt[]{\frac{1}{N} \sum_{i=1}^{n} (\widehat{Y_{i}} - Y_{i}) ^{2}} \end{equation}

Mean Absolute Error (MAE): It is a metric measuring two continuous variables. It takes the sum of the absolute error values because it accurately represents the sum of the error terms. The MAE value is frequently used in regression and time series problems because it is interpretable \cite{13}.
\begin{equation}
MAE = \frac{1}{N} \sum_{i=1}^{n}  \mid (\widehat{Y_{i}} - Y_{i}) \mid   \end{equation}

R square (R$^2$): It is a statistic used to estimate the performance of regression models. The frequency with which the independent variable influences the dependent variable is indicated by this statistic. Additionally, it displays the 0 to 1\% linkage power between the independent and dependent variables.
\begin{equation}
R^{2}=  \frac{({\frac{1}{N} \sum_{i=1}^{n} (Y_{i} -  \underline{Y}) ^{2}})-({\frac{1}{N} \sum_{i=1}^{n} (\widehat{Y_{i}} - Y_{i}) ^{2}})}{{\frac{1}{N} \sum_{i=1}^{n} (Y_{i} -  \underline{Y}) ^{2}}} \end{equation}

\begin{figure}[h]
\centering
\includegraphics[width=\linewidth]{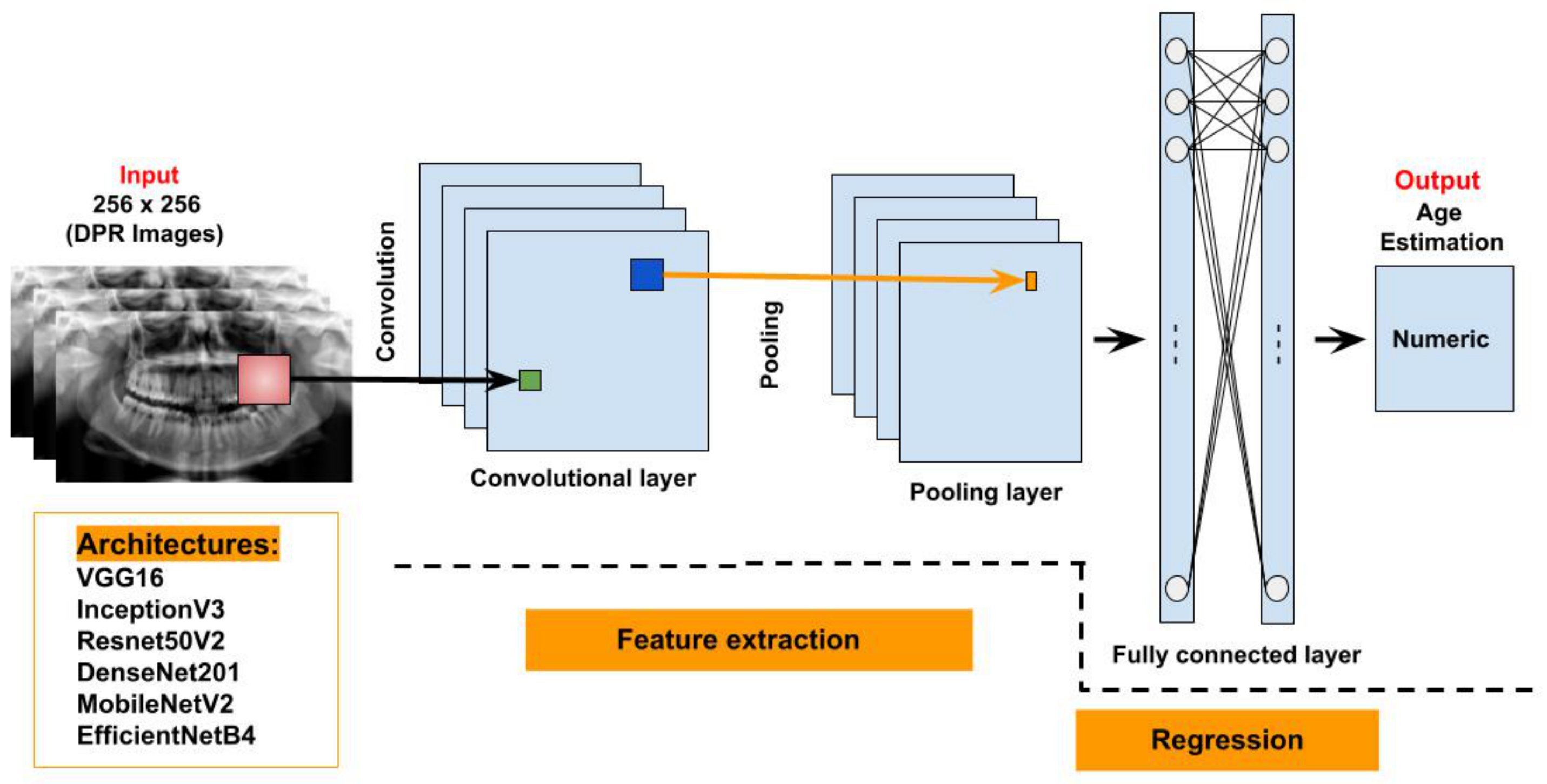}
\caption{Feature Extraction and Regression from DPR Images Based on Deep Learning}
\label{fig:6}
\end{figure}

Here, $(Y_{i})$ refers to the estimated value, while $(\widehat{Y_{i}})$ stands for actual age and $(\underline{Y})$ the true mean age. N denotes the total number of samples. RMSE and MAE are positive values, and these statistical criteria should be smaller. Values close to zero indicate that the forensic age range predictions are reasonably accurate. For the models to be successful, R$^2$ should be close to 1.
Fig. \ref{fig:6} illustrates the general deep learning architecture used for age determination from PDR images. Accordingly, following the loading procedure of the pre-processed DPR images in Fig. \ref{fig:6}, the input feature is extracted in the CNN architecture. Finally, these features are integrated into the regression in the ML layer, and the model is trained for age estimation.

\section{Experimental Results}
\label{sec:4}
This study utilized DPR images for a fully automated forensic age estimation. Deep learning models were used to perform experimental analyses on the provided data sets, and subsequently, the data augmentation technique was applied to increase the model success on images. The data augmentation technique acquires batch images and uses a series of random transformations to each image in the batch. The best values from DPR photographs were determined by analyzing how each function affected the photos in the data triangulation process. Only the noising (normalizing) method of the transfer learning model was applied to the images reserved for the validation and test data sets. For all models, DPR image sizes were scaled to 256 × 256, subjecting the epoch preference to 500 for training. The Adam algorithm was used as the optimization function, taking the default values as reference. However, if the value for validation loss failed to reach the minimum level after every seven epochs, the step size was reduced by 0.8. Finally, the training process ended if the validation loss remained at epoch value 25 (validation loss). The models were not adjusted at all. We took a balanced approach to all models. Since the DPR images were untrained with ImageNet from extensive image databases, parameters were trained entirely for deep learning, and their weights were recalculated accordingly. Of the 1,332 DPR images in the dataset, 962 (85\%), 170 (15\%), and 200 (15\%) were used for training of the selected models, validation, and testing processes, respectively. Table \ref{tab1} lists the performance criteria for the test dataset of the deep learning models applied in the study.
\begin{table}[ht]
\begin{center}
\begin{minipage}{250pt}
\caption{Results according to transfer learning models}\label{tab1}%
\begin{tabular}{@{}ccccc@{}}
\toprule
Model & Total Parameter  & MAE & RMSE & R$^2$\\
\midrule
EfficientNetB4    & 19,5M     & 3.47   & 5.11  & 0.75 \\
ResNet50V2        & 25,6M     & 4.35   & 6.20  & 0.78 \\
DenseNet201       & 20,2M     & 4.22   & 5.99  & 0.80 \\
InceptionV3       & 23,9M     & 3.44   & 5.20  & 0.85 \\
MobileNetV2       & 3,5M      & 3.68   & 5.17  & 0.85 \\
VGG16             & 138,46M   & 10.46  & 12.60 & 0.11 \\
\botrule
\end{tabular}
\end{minipage}
\end{center}
\end{table}

Analysis of Table \ref{tab1} revealed that the InceptionV3 model generated the best performance values in the InceptionV3 model. Parameter models with large quantities seemed to contribute less to the results in training the DPR images. 

Yet, it was also observed that the performance values of both InceptionV3 and MobileNetV2 models were comparable, as the MobileNetV2 model had the smallest parameter among the models selected. The InceptionV3 model was modified to acquire the most optimum values. As a result, several blocks were eliminated from the InceptionV3 model, resulting in the generation of new sub-patterns and analyzing their effects on the outcomes. Table \ref{tab2} lists the performance criteria achieved by these processes. As deduced from Table \ref{tab2}, the InceptionV3Mixed\_04 model produced the best performance values after modifying the InceptionV3 model. Additionally, the InceptionV3Mixed\_06 and InceptionV3Mixed\_07 models generated the most comparable values to these outcomes. The InceptionV3 model had about 24M trainable parameters, while the InceptionV3Mixed\_04 model had 3,68M after the elimination procedure. The experimental studies produced faster and more successful outcomes with almost 6.1 times fewer parameters. 
\begin{table}[ht]
\begin{center}
\begin{minipage}{250pt}
\caption{Performance results of modified InceptionV3 submodels}\label{tab2}%
\begin{tabular}{@{}ccccc@{}}
\toprule
Model & Total Parameter  & MAE & RMSE & R$^2$\\
\midrule
InceptionV3Mixed\_03     & 2,38M     & 3.64   & 5.11  & 0.85 \\
InceptionV3Mixed\_04     & 3,68M     & 3.13   & 4.77  & 0.87 \\
InceptionV3Mixed\_05     & 5,37M     & 3.49   & 5.06  & 0.86 \\
InceptionV3Mixed\_06     & 7,06M     & 3.15   & 4.77  & 0.87 \\
InceptionV3Mixed\_07     & 9,20M     & 3.18   & 4.79  & 0.87 \\
InceptionV3Mixed\_08     & 11,04M    & 3.55   & 5.08  & 0.85 \\
InceptionV3Mixed\_09     & 16,29M    & 3.46   & 4.95  & 0.86 \\
\botrule
\end{tabular}
\end{minipage}
\end{center}
\end{table}

\begin{figure}[ht]
\centering
\includegraphics[width=\linewidth]{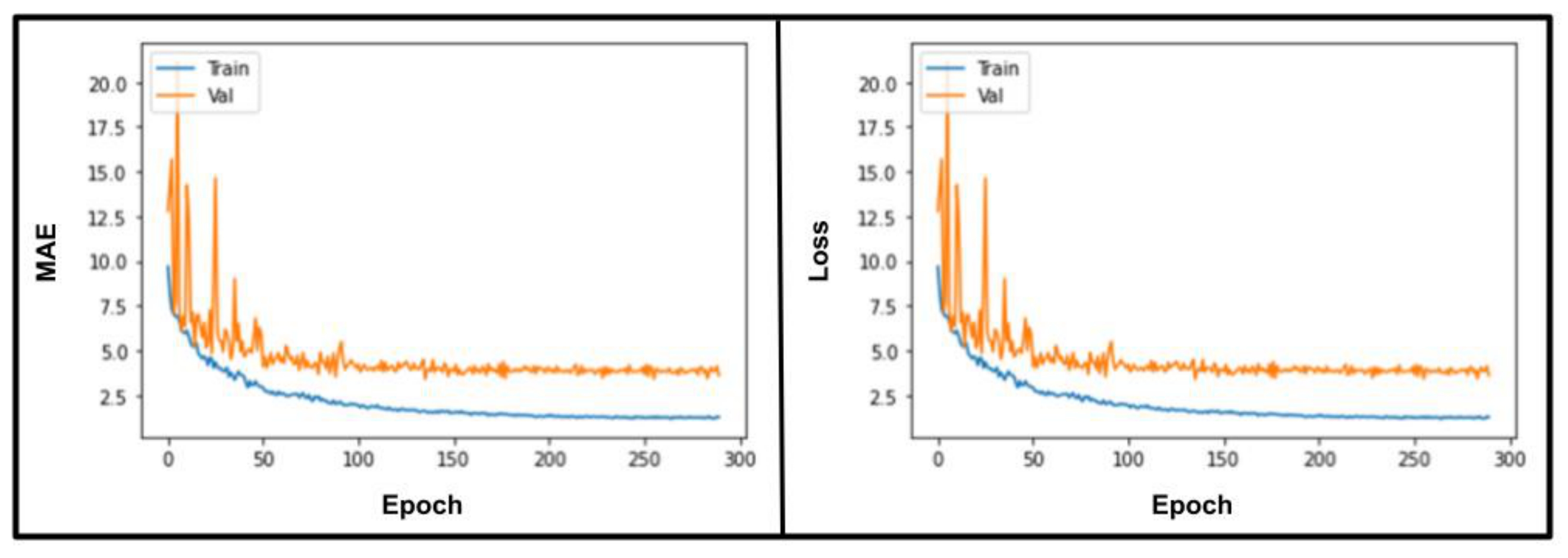}
\caption{MAE and loss graphs of InceptionV3Mixed\_04 model}
\label{fig:7}
\end{figure}

Fig. \ref{fig:7} displays the MAE and loss plots during the training and validation processes of the InceptionV3Mixed\_04 model.

\begin{figure}[ht]
\centering
\includegraphics[width=0.5\linewidth]{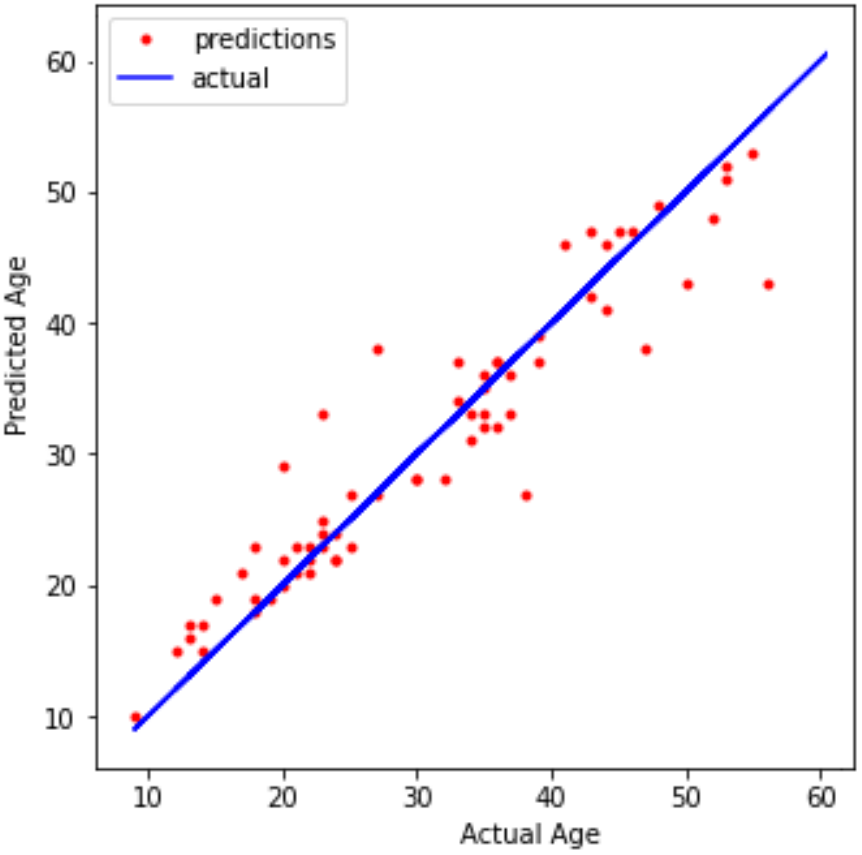}
\caption{Statistical distribution of actual age values and estimated age values}
\label{fig:8}
\end{figure}

Fig. \ref{fig:8} displays the statistical distribution of the DPR image's actual age values and the model's estimated age values from testing with the InceptionV3Mixed\_04 model. Fig. \ref{fig:8} illustrates that the actual and estimated age values of the InceptionV3Mixed\_04 model are reasonably similar to each other. However, the estimation rate was lower for individuals aged over 55. Consequently, the absence or limited image counts in the dataset for ages older than 55 years-old patients and the advanced age degradation in the maxilla and mandible areas could be two potential factors among the reasons. Therefore, it is conceivable to argue that the model's learning capacity for images over 55 years and older was insufficient. 

\begin{figure}[hp]
\centering
\includegraphics[width=0.85\linewidth]{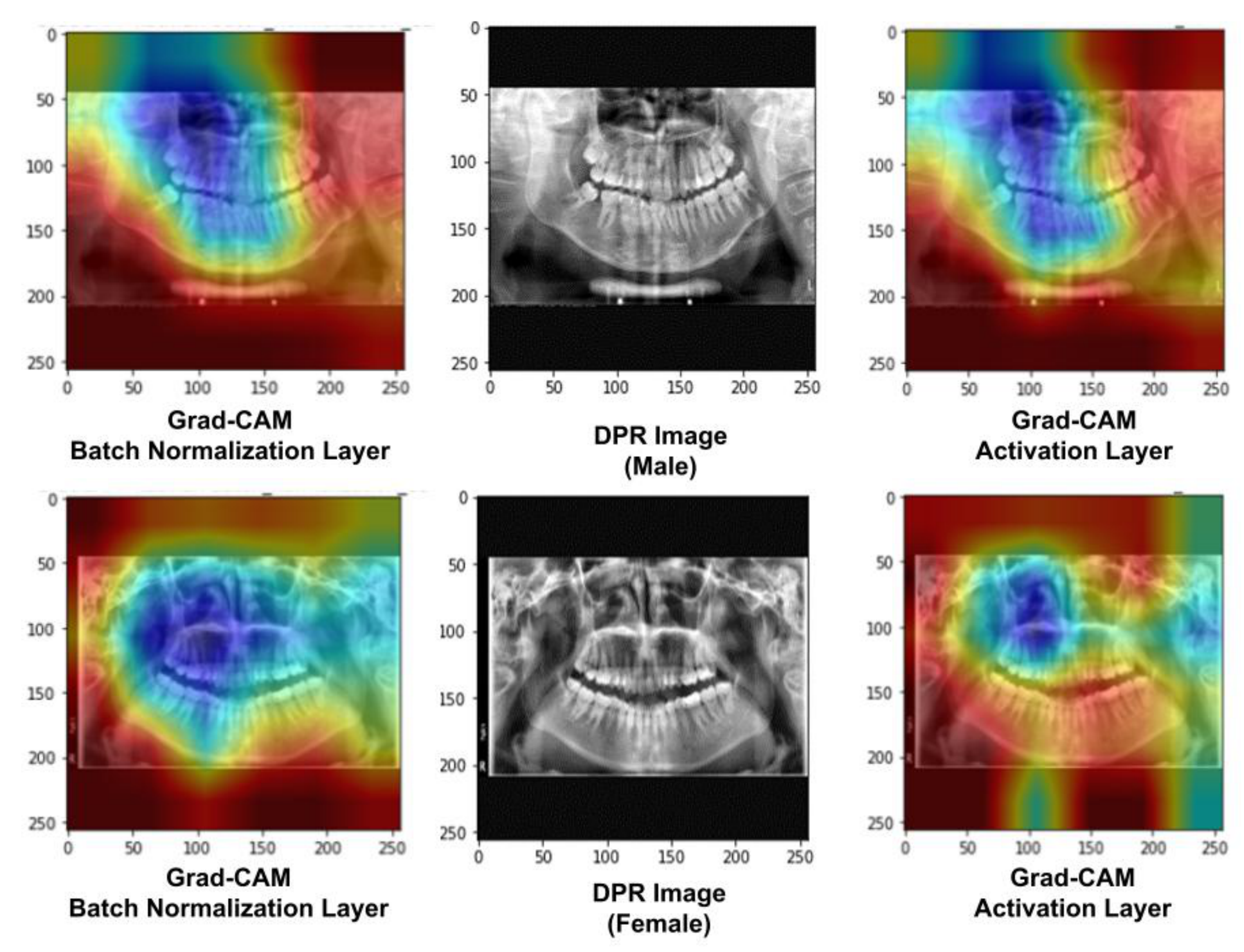}
\caption{Heatmaps of our proposed model on male and female DPR images}
\label{fig:9}
\end{figure}

 A heat map (Grad-CAM) was created to resolve which regions the InceptionV3Mixed\_04 model concentrated. Grad-CAM shows the distinctive areas in the images that a trained model is more likely to detect \cite{34}. Fig. \ref{fig:9} illustrates Grad-CAM heatmaps for the compared female and male DPR tooth images. According to the heatmaps, the InceptionV3Mixed\_04 model focused on the teeth, gingival tissue, and upper jaw (maxilla).

\begin{figure}[ht]
\centering
\includegraphics[width=1\linewidth]{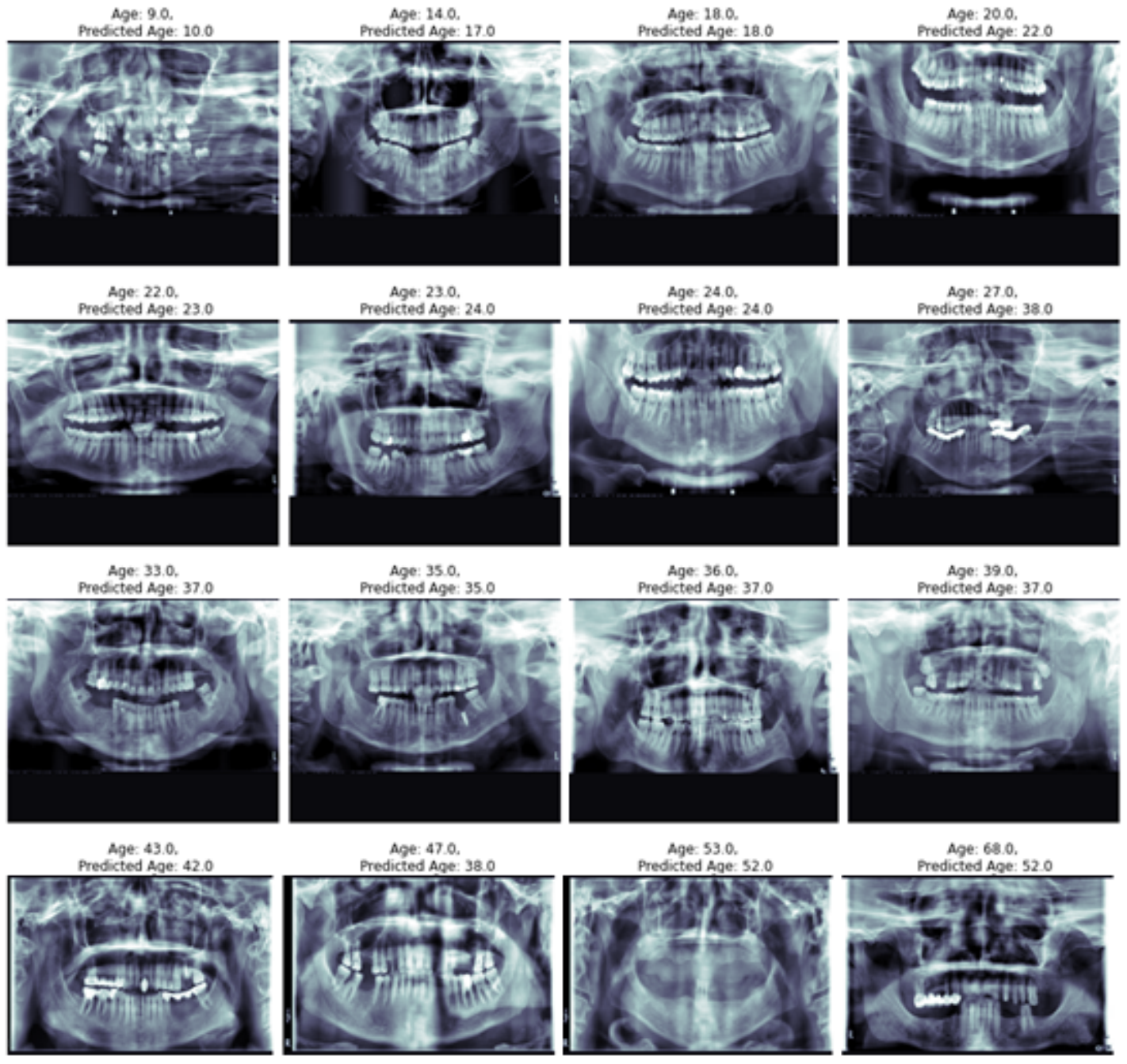}
\caption{Age estimation detection of our proposed model on different DPR images}
\label{fig:10}
\end{figure}

Fig. \ref{fig:10} displays the actual and estimated age values in the various DPR images of the InceptionV3Mixed\_04 model.

Table \ref{tab3} summarizes the literature on Dental Age Estimation using DPR pictures. Since there was no code or DPR dataset for dental age estimation in the literature, we failed to make a rational comparison between our proposed study and the previous studies. 

Analysis of Table \ref{tab3} revealed that studies in the literature typically focused on a limited age range. Kim et al. \cite{22} demarcated the dental ages of individuals between the ages of 2 and 98 by dividing them into age categories. In their study, Asif et al. \cite{23} reported MAE=6.48 - 8.58 in the 16-65 age range (a wide age range). 

This study, however, found the following metric values in the 8-68 age range: MAE=3.13, RMSE=4.77, and R$^2$=0.87. These findings proved that the current study is one of the attempts that has produced the best outcomes in a wide age range on age determination from dental images to date.
\begin{table}[ht]
\begin{center}
\begin{minipage}{250pt}
\caption{Performance results of modified InceptionV3 submodels}\label{tab3}%
\begin{tabular}{@{}ccccc@{}}
\toprule
Model & Province & Age Range  & Dataset & Performance \\
\midrule
 &  &  &   & MAE=2.34 - 4.61  \\
\cite{35} & Poland & 4-15 & 619  &
RMSE=5.58 - 7.49  \\
 &  &   &  & R$^2$=0.92 - 0.96 \\
\midrule
\cite{7} & Croatia & 10-25 & 283  & 
MAE=2.3 - 2.5\\
\midrule
&  &  &   & ACC=89\%  \\
\cite{19} & Malaysian & 4-18 & 426  &  F1 SCORE=71\%  \\
&  &  &   & RECALL=92\%  \\
\midrule
&  &  &   & ACC(14)=95.9\%   \\
\cite{9} & Germany & 5-24 & 10,257  & ACC(16)=95.4\%  \\
&  &  &   & ACC(18)=92.3\%   \\
\midrule
&  &  &   & MAE=2.34 - 4.61   \\
\cite{21} & Germany & 5-25 & 619  &  RMSE=5.58 - 7.49  \\
&  &  &   & R$^2$=0.92 - 0.96   \\
\midrule
&  &  &   & MAE(2-11)=0.8   \\
\cite{22} & Korea & 2-98 & 9,435  & 
MAE(12-18)=1.2  \\
&  &  &   & MAE(19-)=4.4   \\
\midrule
\cite{23} & Malaysian & 16-65 & 300  & MAE=6.48 - 8.58 \\
\midrule
&  &  &   & MAE=3.13   \\
PW\footnotemark[1] & Turkey & 8-68 & 1,332  & 
RMSE=4.77\\
&  &  &   & R$^2$=0.87 \\
\botrule
\end{tabular}
\footnotetext[1]{Proposed Work}
\end{minipage}
\end{center}
\end{table}

\section{Conclusion}
\label{sec:5}
Age determination of individuals is a critical and fundamental element in forensic identification and the work of physicians in administering medications. It is achievable to utilize the traditional age determination methods in forensic medicine after years of education and training procedure of the medical experts. Hitherto, age estimation studies have produced a sizable number of approaches and analyses, each of which touts a variety of uses, criteria for accuracy, and levels of reliability. Nonetheless, age estimations based on dental discoveries are regarded as a depiction of the closely estimated age compared to actual chronological age. Minimizing the average error rates in age estimation is also critical in making the most realistic age determination. This study proposed a deep transfer learning-based modified InceptionV3 model using DPR images. A total of 1,332 DPR images were used for experimental analyses among individuals whose ages ranged from 8 to 68. The modifications in the InceptionV3 model introduced the most optimum results with the InceptionV3Mixed\_04 model. The metrics of the proposed model were MAE=3.13, RMSE=4.77 error rate, and R$^2$=87.2 for accuracy. The current study emphasized the criticalness of modifying transfer learning models to minimize costs and save time rather than developing a new deep learning architecture. As a result, this study developed the InceptionV3Mixed\_04 module with fewer parameters than the InceptionV3 model; consequently, it delivered results faster and estimated ages more precisely. This study also proved that even architectures with fewer parameters of DPR images absent in the ImageNet dataset could still outperform transfer learning models in terms of performance. Generating the heatmap of the InceptionV3Mixed\_04 model uncovered that the right upper jaw (maxilla) and upper teeth (right) were the primary focus of the images. Therefore, we opine that the current study, which aimed to focus on one of the broad age ranges in age estimation, will yield promising outputs for future studies.

\bibliography{sn-bibliography}


\end{document}